\documentclass[aps,pre,twocolumn,superscriptaddress]{revtex4-1}
\usepackage{graphicx,color,epsfig}
\usepackage{amssymb,amsmath,bm}

\begin{document}

\title{Chaos synchronization with coexisting global fields}
\author{O. Alvarez-Llamoza}
\affiliation{Grupo de Simulaci\'on, Modelado, An\'alisis y Accesabilidad, Universidad Católica de Cuenca, Cuenca, Ecuador}
\author{M. G. Cosenza}
\affiliation{School of Physical Sciences \& Nanotechnology, Universidad Yachay Tech, Urcuqu\'i, Ecuador}
\affiliation{Universidad de Los Andes, M\'erida, Venezuela.}

% \date{November 2022}
%\pacs{05.45.-a, 89.75.Kd, 05.45.Xt}

\begin{abstract}
We investigate the phenomenon of chaos synchronization in systems subject
to coexisting autonomous and external global fields by employing a simple model 
of coupled maps. Two states of chaos synchronization are found: (i) complete synchronization, 
where the maps synchronize among  themselves  and  to  the  external  field,  
and  (ii) generalized or internal synchronization, where the maps synchronize 
among themselves but not to the external global field. We show that the stability
conditions for both  states can be achieved for a system of minimum size of two maps. 
We consider local maps possessing robust chaos and characterize the synchronization 
states on the space of parameters of the system. The  state of generalized synchronization
of chaos arises even the drive and the local maps have the same functional form.  
This  behavior is  similar to the process of spontaneous ordering against an external
field found in nonequilibrium systems.
\end{abstract}

\maketitle

\section{Introduction}
A global interaction in a system takes place when all its elements
share a common influence or source of information. 
Global interactions occur in many physical, chemical, biological, social,
and economic systems,
such as parallel electric circuits, coupled oscillators \cite{Kuramoto,Naka}, 
charge density waves \cite{Gruner}, Josephson junction arrays \cite{Wie},
multimode lasers \cite{Wie2}, neural networks, evolution models, ecological systems \cite{Kan1}, 
social networks \cite{Newman}, economic exchange \cite{Yako},
mass media influence \cite{Media1,Media2}, and
cross-cultural interactions \cite{Cross}. 
A variety of phenomena can occur in systems subject to global interactions; for example,
chaos synchronization, dynamical clustering, nontrivial
collective behavior, chaotic itineracy \cite{Kan2,Manrubia}, 
chimera states \cite{Sethia,Yedel}, quorum sensing
\cite{Ojalvo}. These behaviors have been 
investigated in arrays of globally coupled oscillators in diverse experiments \cite{Wang,Monte,Taylor,Tinsley,Scholl}. 
Global interactions can provide  relevant descriptions in networks possessing 
highly interconnected elements or long-range interactions.
Systems with global and local interactions have also been studied \cite{Kan3}.

A global interaction field may consist of an external influence acting on all 
the elements of a system,
as in a driven dynamical system \cite{Parra}; or it may arise from the interactions 
between the elements, such as a mean field \cite{Piko},
in which case we have an autonomous dynamical system. In most situations, 
systems subject to either type of global field have been studied separately. 

In this article, we investigate the dynamics of systems subject to the simultaneous
influence of external and autonomous global interaction fields.
As a simple model for such systems,  we study a network of coupled maps with 
coexisting external and autonomous global fields. 
Specifically, we focus on the important phenomenon of chaos synchronization. 
With this aim, we consider local map units possessing robust chaos dynamics. 
A chaotic attractor is robust if there exists a neighborhood in its
parameter space where windows of periodic orbits are absent \cite{Grebogi}.
Robust chaos is an advantageous property in
applications that require reliable functioning in a chaotic
regime, since the chaotic behavior cannot be
destroyed by arbitrarily small perturbations of the 
parameters. 

In Section~2, we present the model for a system subject to coexisting autonomous
and external global fields. 
We define two types of synchronization states for this system in relation to the external field, 
complete synchronization and generalized synchronization, and characterize them 
through statistical quantities.
In Section~3, we carry out a stability analysis of the synchronization states and 
derive conditions for their stable behavior in terms of the system parameters.
Section~4 contains applications of the model with coexisting global fields for maps
exhibiting robust chaos. The states of chaos synchronization for the system and their
stability boundaries are characterized on the space of parameters expressing the strength 
of the coupling to the global fields. In particular, we show that the state of generalized synchronization of chaos can appear even when the functional forms of the external field 
and the local maps are equal,
a situation that does not occur for this family of maps if only one global field is present.
This behavior represents a collective ordering of the system in a state alternative to that 
of the driving field. Conclusions are given in Section~5.

\section{Coupled map network with autonomous and external global fields}
As a  model of a system of chaotic oscillators with coexisting autonomous and external
global fields, we consider a network of coupled maps in the form

\begin{eqnarray}
y_{t+1} &=& g(y_t) , 
\label{eq1.Modelo}  \\
x_{t+1}^i &=& (1-\epsilon_1-\epsilon_2)f(x_t^i) + \epsilon_1 h_t  + \epsilon_2 g(y_t) , 
\label{eq2.Modelo} \\
 h_t &=& \frac{1}{N}\sum_{j=1}^{N}f(x_t^j) , 
\label{eq3.Modelo}
\end{eqnarray}
where $x_t^i$ represents the states variable of element $i$ $(i=1,2, \ldots,N)$ at
discrete time $t$; 
$N$ is the size of the system;  
$f(x_t^i)$ describes the local chaotic dynamics; 
 $y_{t+1}=g(y_t)$ is an external global field that acts as a homogeneous drive with 
 independent chaotic dynamics; $h_t(x_t^j)$ is an autonomous global field that corresponds 
 to the mean field of the system;
$\epsilon_{1}$ is a parameter measuring the strength of the coupling of the elements
to the mean field $h_t$; and $\epsilon_{2}$ expresses the intensity of the coupling
to the external field. We assume a diffusive form of the coupling for both fields. 

A synchronization state for the system Eqs.~(\ref{eq1.Modelo})-(\ref{eq3.Modelo}) 
occurs when the $N$ elements share the same state; that is, $x_t^i=x_t^j$, $\forall i,j$. 
Then, the mean field becomes $h_t=f(x_t^i)$, $\forall i$. 

Two types of synchronization states can be defined in relation to the external 
global field $g(y_t)$: (i) complete synchronization, where the $N$ elements in 
the system are synchronized among themselves and also to the external driving 
field; i. e., $x_t^i=x_{t}^{j}=y_t$,  $\forall i,j$,  or $h_t=f(x_t^i)=g(y_t)$; and (ii)
internal or generalized synchronization, where the $N$ elements get synchronized 
among themselves 
but not to the external global field; i. e., $x_t^i=x_t^j \neq y_t$,  
$\forall i,j$, or $h_t=f(x_t^i)\neq g(y_t)$.

To characterize the synchronization states
of the system Eqs.~(\ref{eq1.Modelo})-(\ref{eq3.Modelo}) we
calculate the asymptotic time average $\langle\sigma\rangle$ (after discarding transients)
of the instantaneous standard deviation of the
distribution of state variables $\sigma_{t}$, defined as
\begin{equation}
\sigma_{t}=\left[\frac{1}{N}\sum_{i=1}^{N}(x_{t}^{i}-\bar{x}_{t})^{2}\right]^{1/2},
\end{equation}
where 
\begin{equation}
\bar{x}_{t}=\frac{1}{N}\sum_{i=1}^{N}x_t^i.
\end{equation}
 
Additionally, we calculate the asymptotic time average $\langle\delta\rangle$
(after discarding transients) of the
instantaneous difference

\begin{equation}
\delta_t=\left|\bar{x}_t-y_t \right|. 
\end{equation}

Then,
a complete synchronization state  $x_t^i=x_{t}^{j}=y_t$, $\forall i,j$,
where the maps are synchronized to the external global field, is characterized
by the the values $\langle \sigma \rangle=0$ and $\langle \delta \rangle=0$. 
An internal or generalized synchronization state $x_t^i=x_t^j \neq y_t$, $\forall i,j$, 
where the maps are synchronized among themselves but not the external field, 
corresponds to $\langle \sigma \rangle=0$ and $\langle \delta \rangle\neq 0$.

In practice, we set the numerical condition 
$\langle \sigma \rangle < 10^{-7}$ and $\langle \delta \rangle < 10^{-7}$
for the zero values of these statistical quantities to characterize the 
above synchronization states.

\section{Stability analysis of synchronized states}
The system Eqs.~(\ref{eq1.Modelo})-(\ref{eq3.Modelo}) can be written in vector form as 
\begin{equation}
\mathbf{x}_{t+1}=\textbf{M}\mathbf{f}(\mathbf{x}_{t}),
\label{eq.ModeloVectorial}
\end{equation}
where $\mathbf{x}_t$ and  $\mathbf{f}(\mathbf{x}_t)$ are $(N+1)$-dimensional 
state vectors expressed as
\begin{equation}
\mathbf{x}_{t}=\begin{pmatrix} y_t \\ x_{t}^1 \\ x_{t}^2 \\ \vdots \\ x_{t}^N\end{pmatrix}, 
\; \qquad  \mathbf{f}(\mathbf{x}_{t})=\begin{pmatrix}  g(y_t) \\ f(x_{t}^1) \\ f(x_{t}^2) \\ 
\vdots \\ f(x_{t}^N)  \end{pmatrix}, \, 
\end{equation}
and $\textbf{M}$ is the $(N+1)\times(N+1)$ matrix
\begin{equation}
\textbf{M}=(1-\epsilon_1-\epsilon_2)\textbf{I} + \frac{1}{N}\textbf{C},
\end{equation}
where $\mathbf{I}$ is the  $(N+1)\times(N+1)$ identity matrix and $\mathbf{C}$ is the $(N+1)\times(N+1)$ matrix
that represents the coupling to the global fields, given by 
\begin{equation}
\textbf{C}=\begin{pmatrix} 
(\epsilon_1+\epsilon_2) N & 0& \cdots   & 0 \\
\epsilon_{2}N  & \epsilon_1 &\cdots &  \epsilon_1\\
\vdots & \ddots & \vdots & \vdots \\ 
\epsilon_{2}N  & \epsilon_1 & \cdots &  \epsilon_1 \end{pmatrix}.  
\end{equation}

The linear stability
condition for synchronization can be expressed in terms
of the Lyapunov exponents for the system Eq.~(\ref{eq.ModeloVectorial}). 
This requires the knowledge of the $(N+1)$ eigenvalues of matrix $\mathbf{M}$, given by
\begin{equation}
\mu_{k}=(1-\epsilon_1-\epsilon_2) + \frac{1}{N}c_{k}, \qquad  k=1,2,\ldots,N+1,
\end{equation}
where $c_k$ are the eigenvalues of matrix $\mathbf{C}$,
corresponding to $c_1=(\epsilon_1 +\epsilon_2) N$,  $c_2=\epsilon_1 N$, and
$c_k=0$ for $k>2$, which is $(N-1)$--times degenerated. 

Then, the eigenvalues of  matrix $\mathbf{M}$ are
\begin{eqnarray}
\mu_{1} &= & 1, 
\label{e1} \\
\mu_{2} &= & 1-\epsilon_{2}, 
\label{e2} \\
\mu_{k} &= & 1-\epsilon_1-\epsilon_2, \quad   k>2, \\
& & (N-1)-\mbox{times degenerated}.  \nonumber
\label{e3}
\end{eqnarray} 

The eigenvectors of the matrix $\textbf{M}$ satisfying
$\textbf{M}\mathbf{u}_k=\mu_k\mathbf{u}_k$ are also eigenvectors of the matrix $\textbf{C}$, 
and they are given by 

\begin{equation}
\mathbf{u}_1=\begin{pmatrix} 1 \\ 1 \\ \vdots \\  1 \\1 \end{pmatrix}, \, \\ 
\qquad  
\mathbf{u}_2=\begin{pmatrix} 0 \\ 1  \\ \vdots \\ 1 \\ 1 \end{pmatrix}, 
\qquad  
\mathbf{u}_k=\begin{pmatrix} 0 \\ a_1  \\a_2 \\ \vdots\\ a_N \end{pmatrix},  
\end{equation}
where the components $a_k$ of the eigenvectors  $\mathbf{u}_k$, $k>2$, satisfy the condition 
$\sum_{k=1}^N a_{k}=0$, since the eigenvectors $\mathbf{u}_k$ are orthogonal to  
the eigenvectors $\mathbf{u}_1$ and $\mathbf{u}_2$. 

The eigenvectors of the matrix $\mathbf{M}$ constitute a complete basis where the state 
$\mathbf{x}_t$ of the system  Eq.~(\ref{eq.ModeloVectorial}) can be expressed as a
linear combination. In particular, the complete synchronization state of $\mathbf{x}_t$ 
is associated to the eigenvector $\mathbf{u}_1$, while the generalized synchronization state 
is represented by the eigenvector  $\mathbf{u}_2$.

The $(N+1)$ Lyapunov exponents $(\Lambda_1,\Lambda_2,\ldots,\Lambda_{N+1})$ 
of the system Eq.~(\ref{eq.ModeloVectorial}) are defined as
\begin{eqnarray}
&(e^{\Lambda_{1}},e^{\Lambda_{2}},\cdots e^{\Lambda_{N+1}})= & \nonumber\\
& \lim_{T \to \infty}\left(\mbox{magnitude of eigenvalues of} \;
\left|\prod_{t=0}^{T-1}\mathbf{J}(\mathbf{x}_t)\right|\right)^{1/T},  & \nonumber
\label{DefExpLyap}
\end{eqnarray}
where $\textbf{J}$ is the Jacobian matrix of the system Eq.~(\ref{eq.ModeloVectorial}), 
whose components are
\begin{equation}
J_{ij}=\left[(1-\epsilon_1-\epsilon_2)\delta_{ij} + \frac{1}{N}c_{ij}\right]\frac{\partial{[\mathbf{f}(\mathbf{x}_t)]_i}}{\partial{x_j}},
\end{equation}
where $c_{ij}$ are the $ij$-components of matrix $\mathbf{C}$ 
and $[\mathbf{f}(\mathbf{x}_t)]_i$ is the
 $i$-component of vector $\mathbf{f}(\mathbf{x}_t)$. Then, we obtain

\begin{equation}
e^{\Lambda_{k}}=\lim_{T \to \infty}\left|\mu_{k}^{T}\prod_{t=0}^{T-1}f'(x_t^k)\right|^{1/T}, \qquad k=1,\ldots,N+1.
\end{equation}
where $\mu_k$, $k=1,2,\ldots,N+1$, are the eigenvalues of matrix $\mathbf{M}$. 
Substitution of the eigenvalues $\mu_{k}$, gives the Lyapunov
exponents for the system Eq.~(\ref{eq.ModeloVectorial}),
\begin{eqnarray}
\Lambda_{1}&=&\lambda_g, 
\label{Lyap}\\
\Lambda_{2}&=&\ln(1-\epsilon_{2}) +\lim_{T\to \infty}\frac{1}{T}\sum_{t=0}^{T-1}\ln\left|f'(x_t^1)\right|,  \label{Lyap1}\\
\Lambda_{k}&=&\ln(1-\epsilon_1-\epsilon_2) + \lim_{T\to \infty}\frac{1}{T}\sum_{t=0}^{T-1}\ln\left|f'(x_t^k)\right|,\nonumber \\ &  & \hspace{5.1cm} k >2,
\label{Lyap2}
\end{eqnarray}
where $\lambda_{g}$ is the Lyapunov exponent of the driven map $g(y_t)$, 
which is positive since $g(y_t)$ is assumed chaotic.
Note that, in general, the limit terms in Eqs.~(\ref{Lyap1}) 
and (\ref{Lyap2}) depend on $\epsilon_1$ and $\epsilon_2$ since
the iterates $x_t^i$ are obtained from the coupled system 
Eqs.~(\ref{eq1.Modelo})-(\ref{eq3.Modelo}). At synchronization, 
these terms are equal and we denote them by $\lambda_f$.

The stability of the synchronized state is given by the condition
\begin{equation}
e^{\Lambda_k}=\left|\mu_k e^{\lambda_k}\right|<1,
\label{cond-mod}
\end{equation}
where $\lambda_1=\lambda_g$; $\lambda_k=\lambda_f$, for $k>1$. 

Perturbations of the state $\mathbf{x}_t$ along the homogeneous eigenvector
$\mathbf{u}_1=(1,1,\ldots,1)$ do not affect the coherence of the system; 
thus the stability condition corresponding to the eigenvalue $\mu_1$ is 
irrelevant for the complete synchronized state. Then, condition
Eq.~(\ref{cond-mod}) with the next eigenvalue $\mu_2$ provides the 
range of parameter values where the complete synchronized state
$x_t^i=y_t$, $\forall i$, is stable; i. e., 

\begin{equation}
\left|(1-\epsilon_2)e^{\lambda_f}\right|<1 \Rightarrow \quad
1-\frac{1}{e^{\lambda_f}} < \epsilon_2 < 1+\frac{1}{e^{\lambda_f}}.
\label{cs}
\end{equation}
Equivalently, complete synchronization takes place when $\Lambda_2 <0$. 

On the other hand, the internal or generalized synchronization state $\mathbf{x}_t$ 
of the system is proportional to the eigenvector $\mathbf{u}_2=(1,1,\ldots,0)$. 
The stability condition of this state is given by the next degenerate eigenvalue 
$\mu_k$, $k>2$; that is,
\begin{eqnarray}
\left|(1-\epsilon_1-\epsilon_2)e^{\lambda_f}\right|<1 \quad & \nonumber \\
\Rightarrow \; 
 1-\epsilon_2-\frac{1}{e^{\lambda_f}} <\epsilon_1 < 1-\epsilon_2+\frac{1}{e^{\lambda_f}} .&
\label{gs}
\end{eqnarray}
The condition for stable generalized synchronization can 
be also be expressed  as $\Lambda_k <0$.

Because of the eigenvalues Eqs.~(\ref{e1})-(\ref{e3}), the stability 
conditions Eqs.~(\ref{cs}) and (\ref{gs}) can be achieved for any system 
size $N \geq 2$. Equations ~(\ref{cs}) and (\ref{gs}) describe the regions 
on the space of the coupling parameters $(\epsilon_1,\epsilon_2)$ where 
complete and internal synchronization can respectively occur in the system 
with coexisting global fields, Eqs.~(\ref{eq1.Modelo})-(\ref{eq3.Modelo}).

\section{Applications}
We consider the system Eqs.~(\ref{eq1.Modelo})-(\ref{eq3.Modelo}) with local
dynamics described by the tent map
\begin{equation}
f(x_t^i) = \dfrac{r}{2}\left|1-2 \,x_t^i\right|,
\label{eq.Tienda}
\end{equation}
which exhibits robust chaos for $r \in (1,2]$ with $x_t^i \in [0,1]$. 
We fix the local parameter at the value $r=2$ and assume the external 
driving field equal to the local dynamics, i.e., $g=f$. 

\begin{figure}[h]
\centering 
\includegraphics[scale=.75]{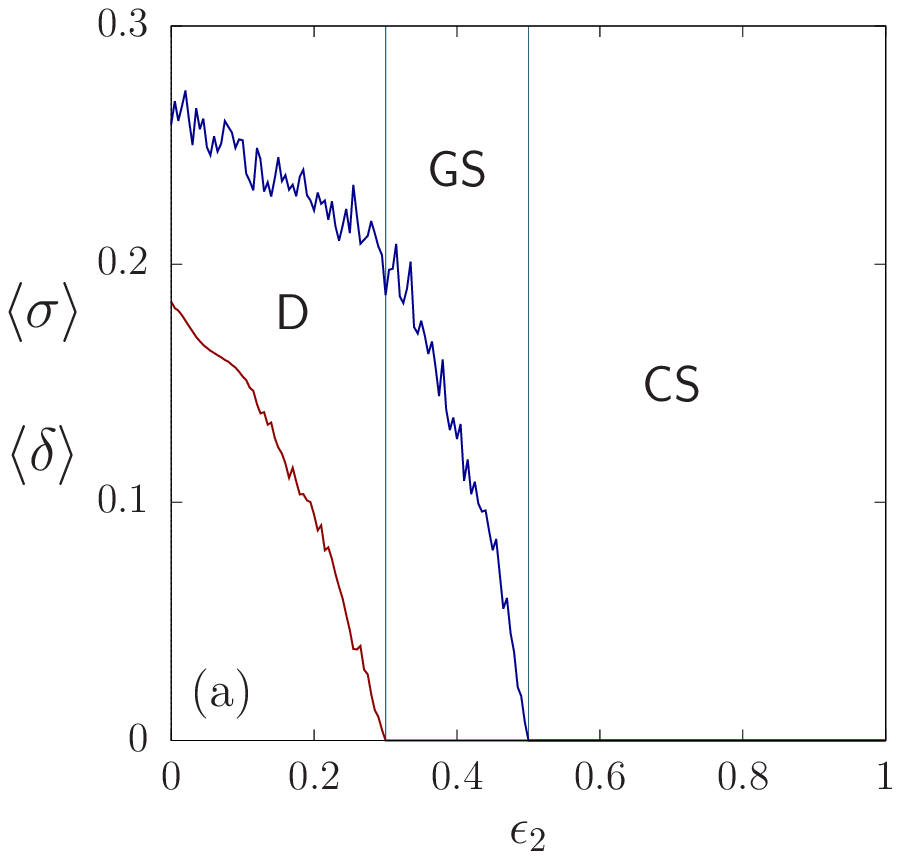}\\
\vspace{0.4cm}
\includegraphics[scale=.75]{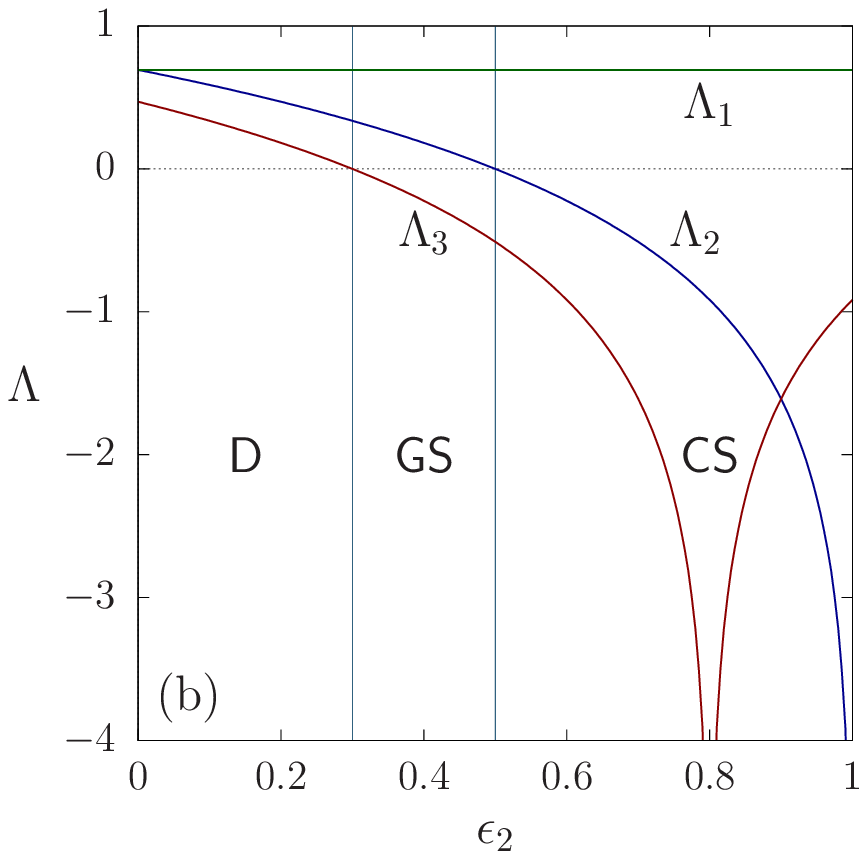}
\caption{(a) Statistical quantities $\langle\sigma\rangle$ (red line) and 
$\langle\delta\rangle$ (blue line) as functions of $\epsilon_2$
for the system Eqs.~(\ref{eq1.Modelo})-(\ref{eq3.Modelo}) with local tent map 
Eq.~(\ref{eq.Tienda}) with $r=2$, and external field $g=f$. Fixed $\epsilon_1=0.2$,
size $N=5000$. For each value of $\epsilon_2$ both quantities are averaged 
over $20000$ iterates after discarding $5000$ transients.
(b) Lyapunov exponents $\Lambda_1=\lambda_g$ (black line), 
$\Lambda_2$ (blue line) and $\Lambda_3$ (red line) as functions of 
$\epsilon_2$ for the system Eqs.~(\ref{eq1.Modelo})-(\ref{eq3.Modelo}) 
with minimum size $N=2$ and $g=f$  with fixed $\epsilon_1=0.2$.
For each value of $\epsilon_2$ the Lyapunov exponents were
calculated with $25000$ iterations after discarding $5000$ transients. 
In this case, $\lambda_f=\lambda_g=\ln 2$.
Labels on both figures indicate D: desynchronized or incoherent state; 
GS: generalized or internal synchronization; CS: complete synchronization.}
\label{Tent}
\end{figure}

Figure~\ref{Tent}(a) shows the statistical quantities $\langle \sigma \rangle$ 
and $\langle \delta \rangle$ that characterize the collective synchronization 
states for this system as functions of the coupling parameter $\epsilon_2$, 
with fixed $\epsilon_1=0.2$. System size is $N=5000$. Labels indicate the
regions of the parameter $\epsilon_2$ where different synchronization states
take place:  D (desynchronized state) where $\langle \sigma \rangle \neq 0$
and $\langle \delta \rangle \neq 0$; SG (generalized or internal  synchronization) 
corresponding to $\langle \sigma \rangle=0$ and $\langle \delta \rangle \neq 0$; 
and  SC (complete synchronization) characterized by  $\langle \sigma \rangle=0$ 
and $\langle \delta \rangle=0$.
Figure~\ref{Tent}(b) shows the Lyapunov exponents $\Lambda_1$, $\Lambda_2$, 
and $\Lambda_3$ as functions of  $\epsilon_2$, with fixed $\epsilon_1=0.2$, 
for a system of minimum size $N=2$, since the stability conditions  for the
synchronized states are satisfied for $N \geq 2$.
The Lyapunov exponent $\Lambda_1=\lambda_g=\ln 2$ is positive. 
The transition of the exponent $\Lambda_3$ from positive to negative values
signals de onset of stable generalized synchronization (GS), 
while $\Lambda_2=0$ indicates the boundary of the complete synchronization state
(CS) that is stable for $\Lambda_2 < 0$. The corresponding boundaries of the 
regions GS and CS coincide exactly in both figures.

\begin{figure}[h]
\centering
\includegraphics[scale=.72]{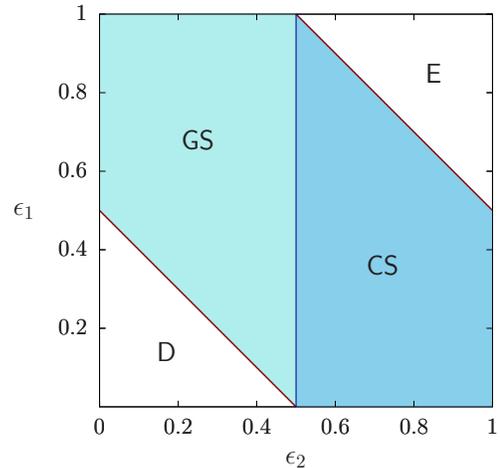}
\caption{Synchronization states for the system 
Eqs.~(\ref{eq1.Modelo})-(\ref{eq3.Modelo}) with  local tent map and external drive $g=f$
on the space of parameters $(\epsilon_1,\epsilon_2)$. Fixed parameters: $r=2, N=5000$. 
Labels indicate the regions where these states can be found:
D: desynchronization; GS: generalized synchronization; CS: complete synchronization;
E: escape. The boundaries determined analytically with Eqs.~(\ref{cs}) and (\ref{gs})
and by the quantities $\langle\sigma\rangle$ and $\langle\delta\rangle$ coincide exactly. 
The region labeled E corresponds to  
coupling parameter values $(\epsilon_1,\epsilon_2)$ for which the state variables 
of the system escape to infinite. The
boundary for region E is given by the upper stability boundary of the generalized 
synchronization state given by Eq.~(\ref{gs}).}  
\label{Curvas_Tienda}
\end{figure}

Figure~\ref{Curvas_Tienda} shows the collective synchronization states of the system Eqs.~(\ref{eq1.Modelo})-(\ref{eq3.Modelo}) with the local tent map and external drive
$g=f$  with $r=2$ on the space of parameters $(\epsilon_1,\epsilon_2)$. 
The regions on this space where the different states occur are indicated by labels. 
The boundaries of the synchronized states are calculated analytically from conditions 
Eqs.~(\ref{cs}) and (\ref{gs}). These boundaries coincide with the criteria for 
the quantities  $\langle\sigma\rangle$ and $\langle\delta\rangle$ characterizing 
each state, as explained above.

\begin{figure}[h]
\centering
\includegraphics[scale=.75]{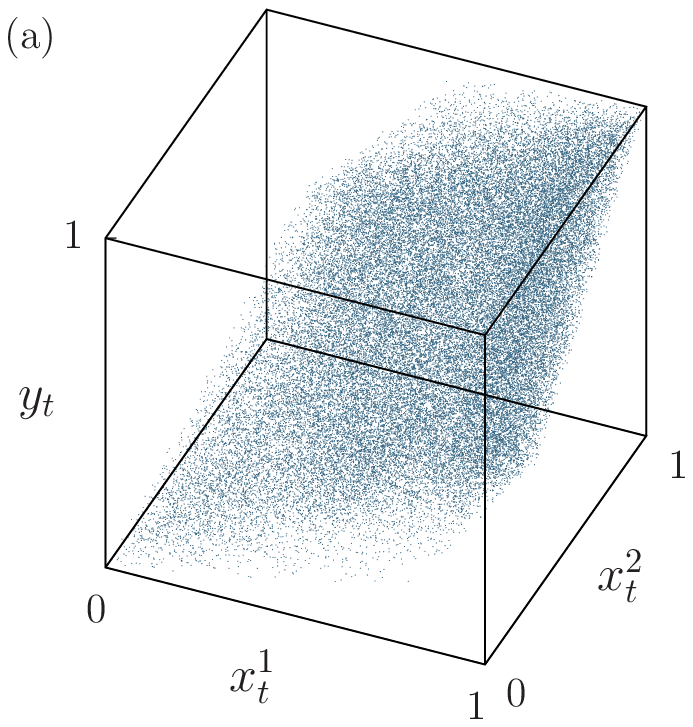} 
\includegraphics[scale=.75]{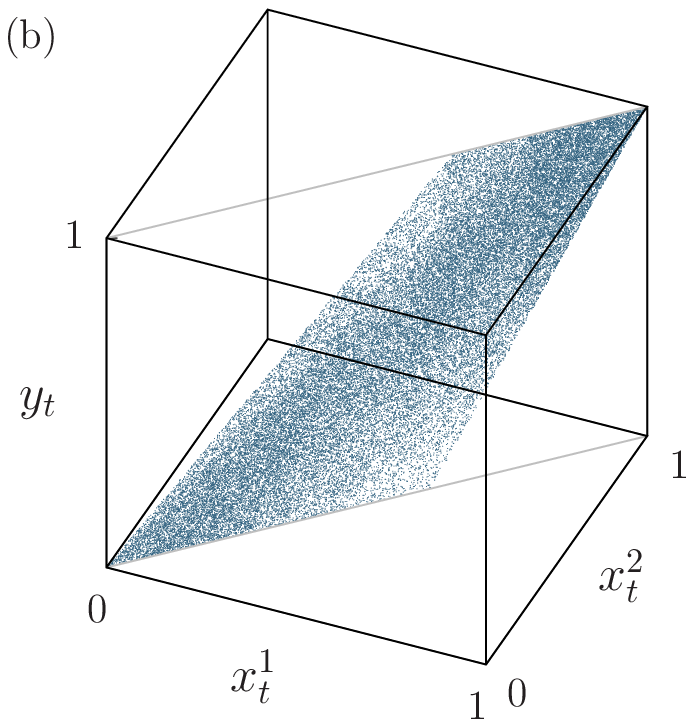} 
\includegraphics[scale=.75]{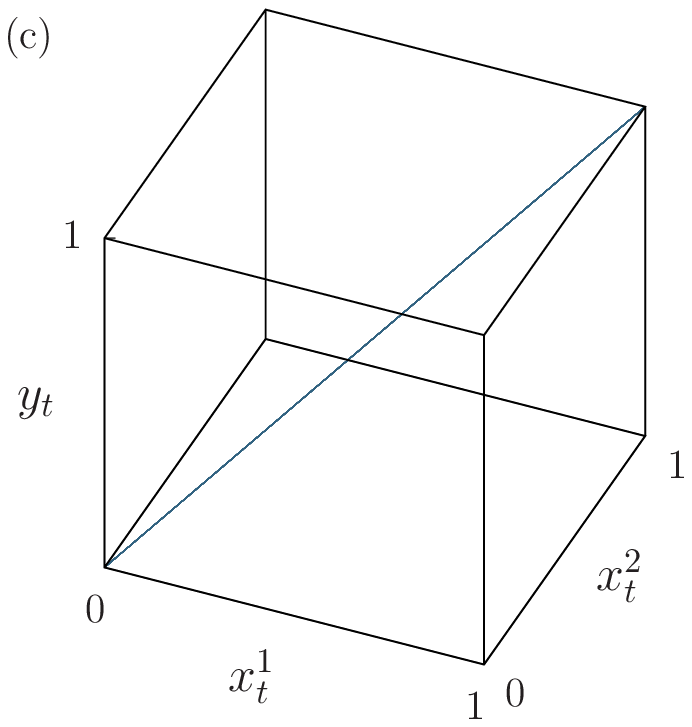}
\caption{Attractors on the three-dimensional phase space of the reduced size system Eqs.~(\ref{eq1.Modelo})-(\ref{eq3.Modelo}) with the local tent map and external drive $g=f$. 
Fixed parameters: $r=2$, $\epsilon_1=0.2$. (a) Desynchronized state (D), $\epsilon_2=0.2$. (b) Generalized synchronization (GS), $\epsilon_2=0.4$. (c) Complete synchronization (CS), 
$\epsilon_2=0.6$.}
\label{Attractor}
\end{figure}

To understand the nature of the collective behaviors, Fig.~\ref{Attractor} 
shows the attractors corresponding to the different synchronization states for
the reduced size system Eqs.~(\ref{eq1.Modelo})-(\ref{eq3.Modelo}), with the local
tent map and external drive $g=f$. The desynchronized state (D) in Fig.~\ref{Attractor}(a)
has all positive Lyapunov exponents and shows no definite structure. In the internal 
synchronization state (GS)  with $\Lambda_1 >0$, $\Lambda_2>0$, and $\Lambda_3 <0$,  
displayed in Fig.~\ref{Attractor}(b), the dynamics collapses onto an attractor 
lying on the plane $x_t^1=x_t^2$. This plane constitutes the synchronization manifold
where $x_t^1=x_t^2=\bar{x}_t$.
Thus, the chaotic attractor on this plane represents a nontrivial functional relation,
different from the identity, between $\bar{x}_t$ and the drive  $y_t$. In general,
for the state of generalized synchronization with $N>2$,  
a chaotic attractor arises between the times series of the mean field $\bar{x}_t$ 
and that of the drive signal $y_t$.
The complete synchronized state (CS), possessing  $\Lambda_1 >0$, $\Lambda_2<0$, $\Lambda_3 <0$, is characterized by  the attractor lying along the diagonal 
line $x_t^1=x_t^2=y_t$, as shown in Fig.~\ref{Attractor}(c). In this situation, 
$\bar{x}_t=y_t$.

The emergence of a chaotic attractor is a characteristic feature of generalized 
synchronization in a drive-response system when the drive function $g$ is different
from the response system function $f$. 
The generalized or internal synchronization state does not arise if only the external drive
$g=f$ acts on the system of tent maps Eqs.~(\ref{eq1.Modelo})-(\ref{eq3.Modelo}), 
i.e., if
$\epsilon_1=0$; nor can it appear with mean field coupling alone. 
The emergence of the GS state in this system
when $g=f$ requires the coexistence of both, the autonomous global field 
and the external drive. 
Thus, we have a situation where the presence of an autonomous global 
interaction allows the synchronization of the maps in a state alternative 
to that of the forcing external field.

As another example, we consider a local chaotic dynamics given the logarithmic map
\begin{equation}
f(x_t^i) = b + \ln\left|x_t^i\right|.
\label{eq.Log}
\end{equation}
This map is unbound and possesses robust chaos, with no windows of periodicity, 
for the parameter interval
 $b\in[-1,1]$ \cite{Jap}.

\begin{figure}[h]
\centering
\includegraphics[scale=.74]{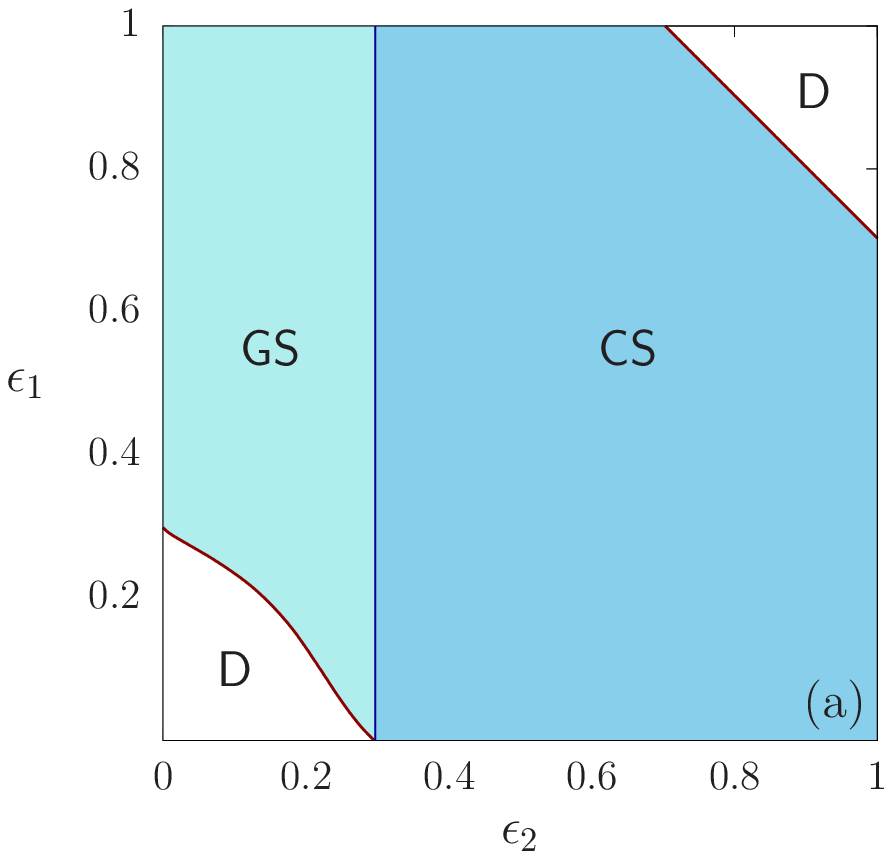} \\
\vspace{0.3cm}
\includegraphics[scale=.74]{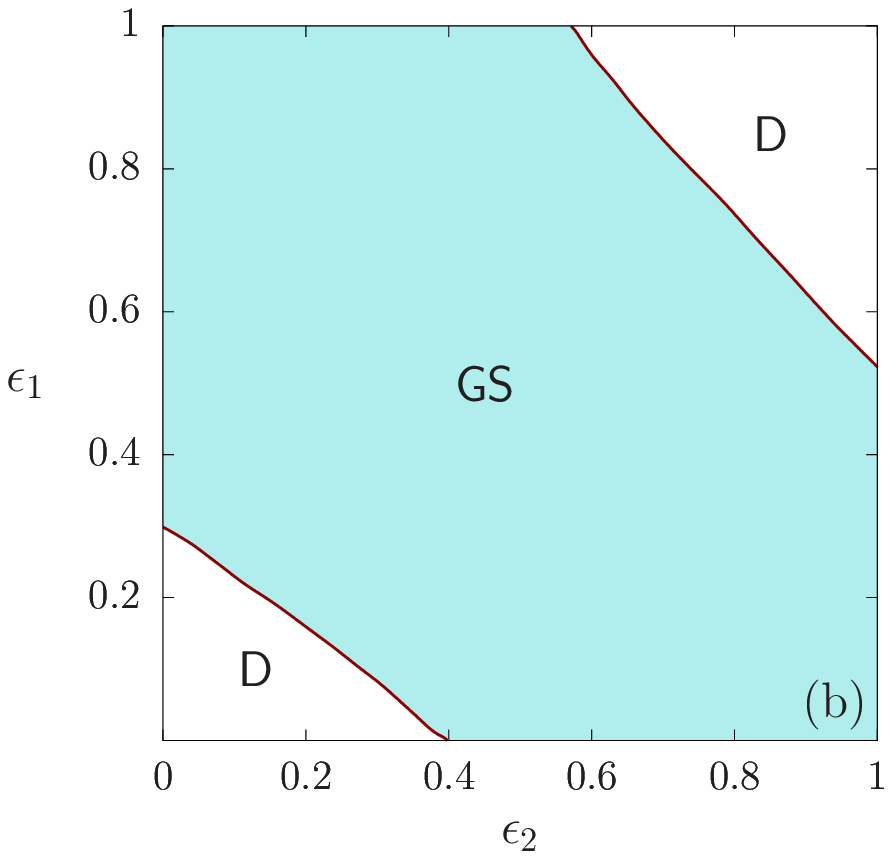} 
\caption{Synchronization states for the system 
Eqs.~(\ref{eq1.Modelo})-(\ref{eq3.Modelo}) with  local logarithmic map Eq.~(\ref{eq.Log})
and $N=5000$ on the space of parameters $(\epsilon_1,\epsilon_2)$. 
(a) External drive function equal to the local dynamics, $g=f=-0.7+\ln|x|$. 
(b) External drive different from the local map, with $g= 0.5+\ln|x|$ and $f=-0.7+\ln|x|$.
Labels indicate the states
D: desynchronization; GS: generalized synchronization; CS: complete synchronization. 
The boundaries determined from conditions Eqs.~(\ref{cs}) (blue line) and (\ref{gs}) 
(red lines) coincide with those obtained  with the quantities $\langle\sigma\rangle$ 
and $\langle\delta\rangle$.}
\label{CL-Loga}
\end{figure}

Figure~\ref{CL-Loga}(a) shows the synchronization states of the system Eqs.~(\ref{eq1.Modelo})-(\ref{eq3.Modelo}) with the local chaotic map  Eq.~(\ref{eq.Log})
and external drive $g=f$  on the plane $(\epsilon_1,\epsilon_2)$. Labels indicate the 
regions where the different synchronization states take place.
The boundaries of the synchronized states are calculated numerically
from Eqs.~(\ref{cs}) and (\ref{gs}). 
The lower boundary of the GS state is calculated 
from Eq.~(\ref{gs}) where $\lambda_f \neq \lambda_g$. For the upper 
boundary of the CS state, the local maps are already synchronized to 
the drive $y_t$ and therefore $\lambda_f =\lambda_g$; thus Eq.~(\ref{gs}) 
gives a straight line on the plane $(\epsilon_1,\epsilon_2)$.
Figure~\ref{CL-Loga}(b) shows the synchronization states corresponding
to an external drive $g \neq f$; only generalized synchronization
(GS) can occur in this case. There is no escape in either situation, 
since the map dynamics is unbounded.

\section{Conclusions}
We have studied a coupled map model for a system subject to coexisting autonomous 
and external global fields. We have investigated the states of chaos synchronization
in this system, consisting of (i) complete synchronization, where the maps synchronize
among themselves and to the external global field, and (ii) generalized or internal 
synchronization, where the maps synchronize among themselves but not to the external field. 
The generalized synchronization state can be described by the appearance of a chaotic 
attractor between the time series of the mean field of the system and the external
driving field. 

We have performed the stability analysis for both synchronization
states and found that the stability conditions can be achieved for a system of
minimum size of two maps subject to a common drive. The equivalence of the dynamics
for a minimum size system is a characteristic feature of systems with global 
interactions \cite{Parra2}. 

By considering local tent maps and logarithmic maps that possess robust chaos dynamics, 
we have focused on the chaotic synchronization behavior of the system. 
We have characterized the synchronization states on the space of the coupling 
parameters by using  the stability conditions of these states as well as
statistical quantities, with complete agreement in all cases. The emergence
of the state of generalized synchronization of chaos, when the drive and the
local maps have the same functional form, requires the presence of both global fields.  
This behavior is similar to the phenomenon of spontaneous ordering against an external
field found in nonequilibrium systems \cite{Media1}. 

Our results suggest that, in addition to chaos synchronization, other collective behaviors either observed or absent
in a system with only one type of global interaction can be modified when
both  autonomous and  external global fields are present. 

\section*{Acknowledgment}
This work was supported by ViceCanciller\'ia de Investigaci\'on e Innovaci\'on, Universidad Yachay Tech, Ecuador. 
% by Corporaci\'on Ecuatoriana para
%el Desarrollo de la Investigaci\'on y Academia (CEDIA).

\end{document}